\documentclass[11pt,twocolumn]{article}
\usepackage[authoryear,sectionbib,sort]{natbib}
\usepackage{url}
\usepackage{amsmath}
\usepackage{abstract}
\begin{document}
\title{Rate of Adaptive Evolution under Blending Inheritance}
\author{Alan R. Rogers\thanks{Dept.\ of Anthropology, 270 S 1400 E Rm
    102, University of Utah, Salt Lake City, UT
    84112. rogers@anthro.utah.edu}} 
\twocolumn[
\maketitle
\begin{onecolabstract}
  In a population of size $N$, adaptive evolution is $2N$ times 
  faster under Mendelian inheritance than the rate implied by
  Victorian theories of heredity and evolution.
\end{onecolabstract}
]
\saythanks

\noindent
{\centering\rule{0.2\linewidth}{0.4pt}\\}
\begin{quote}
  Unfortunately we have no means of determining, according to the
  standard of years, how long a period it takes to modify a species.\\
\mbox{}\hfill Charles \citet[p.~270]{Darwin:OSM-72}
\end{quote}

In the latter half of the 19th century, evolutionary time seemed
limited. Victorian science suggested that the earth was not much older
than 100 million years \citep{Thomson:TRS-23-157, Jackson:AA-01-107}.
Was it really plausible that so much evolution could have happened in
this interval? Evolutionists had no answer. In part, this was because
the question was quantitative---involving rates and intervals---yet
evolutionary theory was at that time merely qualitative.  It was
impossible to say how much time was needed, because (as Darwin laments
in the quote above), no one had a clear understanding of the rate of
evolution.

Well, almost no one. In 1871, A.S.\ Davis \citeyearpar{Davis:N-5-161}
published the first coherent, quantitative model of evolution. But no
one noticed for over a century, until the article was finally
discovered by \citet{Morris:BJH-27-313} and
\citet{Bulmer:BJH-37-281}. In the paragraphs that follow, I ask what
Davis might have concluded about the rate of adaptive evolution, had
he pursued the matter.

Victorians assumed that heredity involved blending the hereditary
material of the parents, and there were at least two theories of
blending \citep{Bulmer:BJH-37-281}. For some, blending was like mixing
paint. For others, it was consistent with an underlying particulate
heredity \citep[p.~12]{Galton:NI-89}.  Following
\citet{Davis:N-5-161}, I assume the paint-pot model of blending
inheritance.

Blending inheritance probably doesn't exist in the real world. Yet it
is important in history, because for half a century it encouraged
skepticism toward natural selection.  Blending implied that selection,
if worked at all, must be very slow, for blending destroys the
variation that makes selection possible.  How slow? The Victorians had
no answer. Yet they could have answered the question had they
tried. This note will provide that answer, by building a model from
Victorian premises. It will not be a modern model, for it will not be
particulate, and it will not incorporate the stochastic effects that
we now see as central to evolution. It will be a model that
A.S.\ Davis might have built, had he put his mind to it.  It will show
that adaptive evolution is far faster under Mendelian inheritance than
the rate implied by Victorian views.

To measure the rate of adaptive evolution, I focus on the rate of
increase in the mean Darwinian fitness of a population. This requires
comment, because it is not clear that mean fitness does increase.
Each adaptive improvement may generate an increase in population size,
which increases competition for resources and leaves the population no
better off than before. We therefore study the marginal effect on mean
fitness, ignoring such compensatory effects as density-dependent
population regulation \citep{Price:AHG-36-129, Frank:TRE-7-92,
  Plutynski:SHP-37-59, Grafen:JRS-52-319}.

Suppose that individuals vary in Darwinian fitness, interpreted as the
probability of surviving from birth to reproductive maturity. As a
cohort matures, its mean fitness rises as selection culls individuals
with low fitness. To measure this effect, let $W$ denote an
individual's absolute fitness. The mean fitness of adults exceeds that
of newborns by\footnote{If $p_W$ is the fraction of newborns with
  fitness $W$, then the corresponding fraction among adults is
  $Wp_W/\bar W$, where $\bar W = \sum W p_W$ is the mean fitness among
  newborns.  The mean fitness of adults is $\sum W^2 p_W/\bar W =
  V/\bar W + \bar W$, where $V$ is the variance in fitness among
  newborns. Subtracting $\bar W$ gives Eqn.~\ref{eq.diffW}, the
  fitness difference between newborns and adults
  \citep{Price:N-227-520}.}
\begin{equation}
\Delta \bar W = V/\bar W
\label{eq.diffW}
\end{equation}
where where $\bar W$ and $V$ are the mean and variance of fitness
among newborns.  This describes changes within a generation, so it
does not depend on the mechanism of inheritance. It works as well with
blending as with Mendelian inheritance.

Let us approximate Eqn.~\ref{eq.diffW} as a differential equation:
\begin{equation}
d \bar W(t) / dt = V(t)/\bar W(t)
\label{eq.dW}
\end{equation}
Under blending inheritance, the variance is halved each generation
\citep{Fisher:GTN-30}:
\begin{equation}
d V(t) / dt = - V(t)/2
\label{eq.dV}
\end{equation}
This equation captures the effect of blending on variance but not that
of selection. Unless selection is very strong, however, its effect
on variance will be negligible compared with that of
blending. Eqn.~\ref{eq.dV} implies that
\begin{equation}
V(t) = V_0 e^{-t/2}
\label{eq.Vt}
\end{equation}
where $V_0$ is the variance in fitness immediately after the
mutation.

Consider the effect of a single new mutation, whose fitness is $1+s$
times $W_0$, the fitness of a normal individual. The initial
population consists of a single mutant and $N-1$ normal
individuals. The initial variance in fitness is
\begin{equation}
V_0 = W_0^2 s^2\left(\frac{1}{N}\right)\left(1 - \frac{1}{N}\right)
\approx W_0^2 s^2/N
\label{eq.V0}
\end{equation}
ignoring terms of order $N^{-2}$.

Equations~\ref{eq.dW}--\ref{eq.V0} imply that mean fitness will evolve
toward an asymptote at which $\bar W_\infty^2 = W_0^2 + 4V_0 = W_0^2(1
+ 4s^2/N)$.  This implies that $\bar W_{\infty}/W_0 \approx 1 +
2s^2/N$.  A single mutation causes a proportional increase,
\begin{equation}
\frac{\bar W_{\infty} - W_0}{W_0} \approx 2s^2/N
\end{equation}
in mean fitness. Using a different argument, Davis derived this result
in 1871 for the special case in which $s=1$.

Let us define mutation rates so as to equalize mutational inputs under
blending and diploid Mendelian inheritance. To this end, assume for
blending that each of the $N$ individuals mutates with probability
$2u$, so that $2Nu$ new mutants arise per generation, each with
selective advantage $s$. The rate of proportional increase in mean
fitness is $2Nu \times 2s^2/N = 4us^2$ under blending inheritance.

Under Mendelian inheritance, the corresponding theory is well
known. Let $u$ represent the probability per generation that a single
gene mutates to a new allele, whose heterozygous carriers have fitness
$1+s$ times that of normal individuals. There are on average $2Nu$
such mutations per generation, of which a fraction $2s$ is eventually
fixed \citep{Haldane:MPC-23-838}. If gene effects are additive, so
that homozygous mutants have fitness $1+2s$, then each fixed mutation
increases mean fitness by $2s$.  The product, $8Nus^2$, of these
quantities, is the rate of proportional increase in mean fitness under
Mendelian inheritance.

This is $2N$ times the rate under blending inheritance. In a
population of 10,000 individuals, adaptive evolution would be 20,000
times slower under blending than under Mendelian inheritance, given
equal mutational inputs.

Late in the 19th century, two misconceptions undermined the debate
about evolutionary time. One of these---the age of the earth---has
been widely discussed. Yet by comparison, its effect was
minor. Victorians underestimated the age of the earth by 1 or 2 orders
of magnitude. The error implied by their theory of heredity was much
larger.

Evolutionists have long known that Mendelian inheritance solved a
fundemental evolutionary problem---the maintenance of variation---and
also that evolutionary rates depend on variance. We have not, however,
fully understood the overwhelming advantage in speed that Mendelism
gave to evolution.

\paragraph{Acknowledgements} I am grateful for comments from Guillaume
Achaz, Thomas Bataillon, Elizabeth Cashdan, Stephen Downes, Steven
Gaulin, and Sergey Kryazhimskiy.

\bibliographystyle{amnatnat}
\bibliography{defs,arr,molrec,gs,dating}
\end{document}